\documentclass{PoS}

\title{Threshold resummation beyond leading eikonal level}

\ShortTitle{Threshold resummation beyond leading eikonal level}

\author{\speaker{Georges Grunberg}%
        \\
   Centre de Physique Th\'eorique,  \'Ecole
Polytechnique,\\
        91128 Palaiseau Cedex, France\\
       E-mail: \email{georges.grunberg@cpht.polytechnique.fr}}

\abstract{The modified evolution equation for parton distributions of Dokshitzer, Marchesini and Salam is extended to non-singlet Deep Inelastic Scattering coefficient functions and the physical evolution kernels which govern their scaling violation. Considering the $x\rightarrow 1$ limit, it is found that the leading  next-to-eikonal   logarithmic contributions to the momentum space physical kernels at any loop order can be expressed in term of the one loop cusp anomalous dimension, a result which can  presumably be extended to all orders in $(1-x)$. Similar results  hold for fragmentation functions in semi-inclusive $e^{+}e^{-}$ annihilation. The method does not work for subleading next-to-eikonal logarithms, but, in the special case of the $F_1$ and $F_T$ structure and fragmentation functions, there are hints of the possible existence of an underlying  Gribov-Lipatov like relation.}

\FullConference{XVIII International Workshop on Deep-Inelastic Scattering and Related Subjects\\
                April 19 -23, 2010\\
                Convitto della Calza, Firenze, Italy}

\begin{document}

\section{Threshold resummation in physical evolution kernels}
Consider  a generic  deep inelastic scattering (DIS) {\em non-singlet} structure function ${\cal F}(x,Q^2)=\{2F_1(x,Q^2),F_2(x,Q^2)/x\}$
at large $Q^2>>\Lambda^2$.
We shall be interested in the {elastic limit} $x\rightarrow 1$  where the final state mass 
$W^2\sim (1-x)Q^2<<Q^2$. 
In this limit, large threshold $\ln (1-x)$ logarithms appear. Their resummation is by now standard 
\cite{Sterman:1986aj,Catani:1989ne}, but usually performed in moment space. However,
the result can also be expressed {\em analytically in  momentum space} at the level of so-called  ``physical  evolution kernels'' which account for the
  {\em physical} scaling violation:
  
  \begin{equation}\frac{\partial {\cal F}(x,Q^2)}{\partial \ln Q^2}\,=\,\int_x^1 \frac{dz}{z}\,  K(z,a_s(Q^2))\,{\cal F}(x/z,Q^2)\equiv K \otimes {\cal F}\label{physicalkernel}\ ,\end{equation}
where the ``physical  evolution kernel''  $K(x,a_s)$ ($a_s=\alpha_s/4\pi$ is the $\overline{MS}$ coupling) embodies {\em all} the {\em perturbative} information about ${\cal F}$.
For $x\rightarrow 1$ threshold resummation yields \cite{Gardi:2007ma}:

\begin{equation}K(x,a_s(Q^2))\sim\left[\frac{{\cal J}
\left((1-x)Q^2\right)}{1-x}\right]_{+} + B_{\delta}^{DIS}(a_s(Q^2))\, \delta(1-x)\label{large-x-eikonal}\ ,\end{equation}
where ${\cal J}(Q^2)$ is a  ``physical  anomalous dimension'' (a renormalization scheme invariant quantity), related to the standard ``cusp''  $A(a_s)=\sum_{i=1}^\infty
A_ia_s^{i}$ and final state ``jet function'' $B(a_s)=
\sum_{i=1}^\infty B_i a_s^{i}$ anomalous dimensions by:

\begin{equation} \label{standard-J-coupling}
{\cal J}(Q^2)=A\left(a_s(Q^2)\right)+\beta\left(a_s(Q^2)\right)\frac{dB\left(a_s(Q^2)\right)}{da_s}\equiv\sum_{i=1}^\infty
j_i  a_s^{i}(Q^2)
\ .\end{equation}
The renormalization group invariance of ${\cal J}(Q^2)$ yields the standard relation:

\begin{eqnarray}{\cal J}\left((1-x)Q^2\right)&=&j_{1}\ a_s+a_s^2[-j_{1}\beta_0 L_x+j_{2}]\label{j-expand}\\
&+ &a_s^3[j_{1}\beta_0^2 L_x^2-(j_{1}\beta_1+2 j_{2} \beta_0) L_x+j_{3}]+{\cal O}(a_s^4)\nonumber\ ,
\end{eqnarray}
where  $L_x\equiv \ln(1-x)$ and $a_s=a_s(Q^2)$, from which the structure of {\em all} the eikonal logarithms in $K(x,a_s(Q^2))$, which can be absorbed into the {\em single} scale $(1-x)Q^2$, can thus be derived. 

However, no analogous result holds at the next-to-eikonal level (except \cite{Grunberg:2007nc} at large-$\beta_0$). Indeed, expanding

\begin{equation}
 K(x,a_s)=K_0(x) a_s+K_1(x) a_s^2+K_2(x) a_s^3+{\cal O}(a_s^4)\label{K-expand}\ ,\end{equation}
the $K_i$'s can be determined as combinations of splitting and coefficient functions. One gets:

\begin{equation}\label{P0}K_0(x)= P_0(x)=k_{10}\ p_{qq}(x)+\Delta_1\, \delta(1-x)\ ,\end{equation} 
with $k_{10}=A_1$ and $ p_{qq}(x)=\frac{x}{1-x}+\frac{1}{2}(1-x)$.
Moreover for $x\rightarrow 1$ one finds 
\cite{Grunberg:2009yi,Moch:2009hr}, barring delta function contributions:

\begin{eqnarray}
\label{logstructure}K_1(x)&=&\frac{x}{1-x}(k_{21}L_x+k_{20})+ (h_{21}L_x+h_{20})+{\cal O}((1-x)L_x)\\
K_2(x)&=& \frac{x}{1-x}(k_{32}L_x^2+k_{31}L_x+k_{30})+ (h_{32}L_x^2+h_{31}L_x+h_{30})+{\cal O}((1-x)L_x^2)\nonumber \ .\end{eqnarray}
Despite the similar logarithmic structure,  the {\em next-to-eikonal} logarithms $h_{ij}$ {\em cannot}   \cite{Grunberg:2009yi} be obtained from a standard renormalization group resummation analogous to the one used (eq.(\ref{j-expand})) for
the {\em eikonal} logarithms $ k_{ij}$.

\section{An alternative approach: the modified physical kernel}
Instead, consider \cite{Grunberg:2009vs} a  modified physical evolution equation, similar to the one used in \cite{Dokshitzer:2005bf} (see also \cite{Basso:2006nk}) for parton distributions:

\begin{equation}\label{Knew}
\frac{\partial {\cal F}(x,Q^2)}{\partial \ln Q^2}\,=\,\int_x^1 \frac{dz}{z}\, K(z,a_s(Q^2),\lambda)\,{\cal F}(x/z,Q^2/ z^{\lambda})\ ,
\end{equation}
where the arbitrary parameter $\lambda$ shall  be set to $1$ at the end.
Expanding ${\cal F}(y,Q^2/ z^{\lambda})$ around $z=1$, one can relate $K(x,a_s,\lambda)$  to $K(x,a_s)$:

\begin{equation}
\label{Knew-Kold} K(x,a_s, \lambda)=K(x,a_s)+\lambda [\ln x\  K(x,a_s, \lambda)]\otimes K(x,a_s)+...\ . \end{equation}
Solving perturbatively, one finds that for $x\rightarrow 1$ the corresponding expansion coefficients $K_i(x,\lambda)$ satisfy the analogue of eq.(\ref{logstructure}), with the {\em same} coefficients $k_{ji}$'s of the eikonal logarithms, but with the coefficients of the {\em leading} next-to-eikonal logarithms given by:

\begin{eqnarray}
\label{hji} h_{21}(\lambda)&=&h_{21}-\lambda k_{10}^2\\
h_{32}(\lambda)&=&h_{32}-\lambda\frac{3}{2}k_{21}k_{10}\nonumber\ .\end{eqnarray}
Setting now $\lambda=1$, one observes that both $h_{21}(\lambda=1)$ and $h_{32}(\lambda=1)$ {\em vanish}, which means that  $h_{21}=k_{10}^2=A_1^2=16C_F^2$ and $h_{32}=\frac{3}{2}k_{21}k_{10}=-\frac{3}{2}\beta_0 A_1^2=-24 \beta_0 C_F^2 $, which agree with the exact results in \cite{Grunberg:2009yi,Moch:2009hr}. It should be stressed that, whereas  $h_{21}$ is contributed by the two loop splitting function alone (and thus one simply recovers in this case the result of \cite{Dokshitzer:2005bf}), $h_{32}$ is instead contributed {\em only} by the one and two loop coefficient functions, which represents a new result. Similar results are obtained for the coefficients $h_{ji}$ ($j=i+1$) of the  {\em leading} next-to-eikonal logarithms at any loop order, which can all be expressed in term of the one loop cusp anomalous dimension  {\em assuming} the  corresponding $h_{ji}(\lambda)$  vanish for $\lambda=1$. In particular, one predicts $h_{43}=\frac{4}{3} k_{10}k_{32}+\frac{1}{2}k_{21}^2=\frac{11}{6}\beta_0^2 A_1^2=\frac{88}{3} \beta_0^2 C_F^2 $, which is correct \cite{Grunberg:2009yi,Moch:2009hr}, and $h_{54}=\frac{5}{4} k_{10}k_{43}+\frac{5}{6}k_{21}k_{32}=-\frac{25}{12}\beta_0^3 A_1^2=-\frac{100}{3} \beta_0^3 C_F^2$, which remains to be checked.

\noindent Similar results are obtained for the coefficients $f_{ji}$ ($j=i+1$) of the {\em leading} next-to-next-to eikonal logarithms, defined by:
 
\begin{equation}
\label{LL}\left. K_i(x)\right\vert_{\rm LL}=L_x^i[ p_{qq}(x)\ k_{j  i}+h_{j i}+(1-x)f_{j  i}+{\cal O}((1-x)^2)]\ , \end{equation}
where the {\em full} one loop prefactor $p_{qq}(x)$ should be used in the leading term to define the  $f_{ji}$'s. The corresponding $f_{j  i}(\lambda)$ coefficients in $K_i(x,\lambda)$ are given by:

\begin{eqnarray}\label{f-lambda}
f_{21}(\lambda)&=&f_{21} +\lambda\frac{1}{2}k_{10}^2\\
f_{32}(\lambda)&=&f_{32}-\lambda(-\frac{3}{4}k_{10}k_{21}+k_{10}h_{21})+\lambda^2\frac{1}{2}k_{10}^3\nonumber\\
f_{43}(\lambda)&=&f_{43}-\lambda\Big(-\frac{2}{3}k_{10}k_{32}+\frac{1}{2}(h_{21}-\frac{1}{2}k_{21})k_{21}+k_{10}h_{32}\Big)+\lambda^2 k_{10}^2 k_{21}\nonumber\ ,\end{eqnarray}
where   one  notes the presence of contributions {\em quadratic} in $\lambda$. Assuming  the  $f_{j  i}(\lambda)$'s vanish for $\lambda=1$, the resulting predictions for the $f_{ji}$'s ($j=i+1$) are again found to agree with the exact results of \cite{Moch:2009hr}.

\section{Fragmentation functions}
Similar results hold for physical evolution kernels associated to fragmentation functions in semi-inclusive $e^{+}e^{-}$ annihilation (SIA), provided one sets $\lambda=-1$ in the modified  evolution equation:

\begin{equation}
\label{SIAkernel}
\frac{\partial {\cal F}_{SIA}(x,Q^2)}{\partial \ln Q^2}\,=\,\int_x^1 \frac{dz}{z}\,K_{SIA}(z,a_s(Q^2),\lambda)\,{\cal F}_{SIA}(x/z,Q^2/z^{\lambda})\ ,
\end{equation}
where ${\cal F}_{SIA}=\{{\cal F}_{T}, {\cal F}_{T+L}\}$ denotes a generic {\em non-singlet} fragmentation function (I use the notation of \cite{Moch:2009hr}).
At the {\em leading} eikonal level, threshold resummation \cite{Cacciari:2001cw}
can be summarized in the standard SIA physical evolution kernel by:

\begin{equation}
\label{SIAkernel-largex} K_{SIA}(x,a_s(Q^2))\sim \left[\frac{{\cal J}
\left( (1-x)Q^2\right)}{1-x}\right]_{+} + B_{\delta}^{SIA}(a_s(Q^2))\, \delta(1-x)\ ,\end{equation}
where the ``physical anomalous dimension'' ${\cal J}(Q^2)$ (hence the $k_{ji}$'s) are the {\em same} for DIS and SIA, as follows from the results in \cite{Moch:2009my}. Assuming the {\em leading} threshold logarithms {\em vanish} beyond the leading  eikonal level in the {\em modified} SIA evolution kernel for  $\lambda=-1$, and setting $\lambda=-1$ in eq.(\ref{hji}) and (\ref{f-lambda}), one derives predictions for $h_{ji}^{SIA}$ and $f_{ji}^{SIA}$ ($j=i+1$) which again agree with the exact results of  \cite{Moch:2009hr}. In particular, one finds that $h_{ji}^{SIA}=-h_{ji}$.

\section{Subleading next-to-eikonal logarithms}
The previous approach {\em does not} work for {\em subleading} next-to-eikonal logarithms, namely the latter do not vanish in the modified physical evolution kernels for $\lambda=\pm 1$.
The following  facts are nevertheless worth quoting:
\begin{itemize}
\item[$ \bullet$] At {\em large} $\beta_0$, we have a generalization \cite{Grunberg:2007nc} of the leading eikonal {\em single scale} ansatz (which takes care of {\em all} subleading logarithms)  to {\em any} eikonal order:

\begin{eqnarray}
\label{large-beta}
\left. K(x, Q^2)\right\vert_{\rm large \,\,\beta_0}&=&\left[\frac{x}{1-x}\left.{\cal J}
(W^2)\right\vert_{\rm large \,\,\beta_0}\right]_{+} +\left(\delta(1-x) term\right)\\
& &+\left.{\cal J}_0( W^2)\right\vert_{\rm large \,\,\beta_0}+(1-x)
\left.{\cal J}_1( W^2)\right\vert_{\rm large \,\,\beta_0}+...
\nonumber\end{eqnarray}
where $W^2=(1-x) Q^2$, and the ${\cal J}_i$'s ({\em except} the leading eikonal one) are structure function dependent .
A similar result holds for $\left. K_{SIA}(x, Q^2)\right\vert_{\rm large \,\,\beta_0}$.
\item[$ \bullet$] There are remarkable  relations between the {\em momentum space next-to-leading}  threshold  logarithms of the (DIS)  $ F_1$ and  the corresponding (SIA)  $F_T$ transverse fragmentation function physical evolution kernels at the {\em next-to-eikonal} level. Namely, using the moment space results of 
\cite{Moch:2009hr},
one can  derive the following {\em momentum space} relations: 

1) At two loop for the ${\cal O}(L_x^0)$  next-to-eikonal constant term:

\begin{eqnarray}
\label{h-NLL-2loop}
 h_{20}^{(F_1)}&=&\left. h_{20}^{(F_1)}\right\vert_{\rm large \,\,\beta_0}+\Delta h_{20}\\
 h_{20}^{(F_T)}&=&\left. h_{20}^{(F_T)}\right\vert_{\rm large \,\,\beta_0} -\Delta h_{20}\nonumber\ ,\end{eqnarray}
with
$\left. h_{20}^{(F_1)}\right\vert_{\rm large \,\,\beta_0}=-11\beta_0 C_F$, 
$\left. h_{20}^{(F_T)}\right\vert_{\rm large \,\,\beta_0}=7\beta_0 C_F$,
and
$\Delta h_{20}=A_1 \Delta_1=12 C_F^2$.

2) At three loop for the {\em single} ${\cal O}(L_x)$ next-to-eikonal logarithms:

\begin{eqnarray}
\label{h-NLL-3loop}
 h_{31}^{(F_1)}&=&\left. h_{31}^{(F_1)}\right\vert_{\rm large \,\,\beta_0} +\Delta h_{31}\\
 h_{31}^{(F_T)}&=&\left. h_{31}^{(F_T)}\right\vert_{\rm large \,\,\beta_0} -\Delta h_{31}\nonumber\ ,\end{eqnarray}
with
$\left. h_{31}^{(F_1)}\right\vert_{\rm large \,\,\beta_0}=-2\beta_0\, \left. h_{20}^{(F_1)}\right\vert_{\rm large \,\,\beta_0}=22 C_F \beta_0^2$,
$\left. h_{31}^{(F_T)}\right\vert_{\rm large \,\,\beta_0}=-2\beta_0\, \left. h_{20}^{(F_T)}\right\vert_{\rm large \,\,\beta_0}=-14 C_F \beta_0^2$,
and:

\begin{equation}\Delta h_{31}=2A_1 A_2-20\beta_0 C_F C_A+20\beta_0 C_F^2\ .\end{equation}

3) At four loop for the {\em double} ${\cal O}(L_x^2)$ next-to-eikonal logarithms:

\begin{eqnarray}
\label{h-NLL-4loop}
 h_{42}^{(F_1)}&=&\left. h_{42}^{(F_1)}\right\vert_{\rm large \,\,\beta_0} +\Delta h_{42}\\
 h_{42}^{(F_T)}&=&\left. h_{42}^{(F_T)}\right\vert_{\rm large \,\,\beta_0} -\Delta h_{42}\nonumber\ ,\end{eqnarray}
with
$
\left. h_{42}^{(F_1)}\right\vert_{\rm large \,\,\beta_0}=3\beta_0^2\, \left. h_{20}^{(F_1)}\right\vert_{\rm large \,\,\beta_0}=-33 C_F \beta_0^3$,
$\left. h_{42}^{(F_T)}\right\vert_{\rm large \,\,\beta_0}=3\beta_0^2\, \left. h_{20}^{(F_T)}\right\vert_{\rm large \,\,\beta_0}=21 C_F \beta_0^3$, and:

\begin{equation}
 \Delta h_{42}= -24\beta_1 C_F^2+45\beta_0^2 C_F C_A-178\beta_0^2 C_F^2
 -(47-10\zeta_2)\beta_0 C_F C_A^2-(60-140\zeta_2)\beta_0 C_F^2 C_A-16\beta_0 C_F^3\  .\end{equation}

The large-$\beta_0$ parts are consistent with eq.(\ref{large-beta}), while the remaining $\pm\Delta h_{ij}$ corrections are suggestive of an underlying (yet to be discovered) Gribov-Lipatov like relation \cite{Gribov:1972ri}.

 \item[$ \bullet$] {\em No such relations} exist between the DIS $F_2$  structure function and the corresponding total angle-integrated $F_{T+L}$ fragmentation function. This fact suggests to focus instead on the {\em momentum space} physical evolution kernels of the {\em longitudinal} structure \cite{Moch:2009mu,Grunberg:2009am} and fragmentation functions. 
Indeed,  some observations in \cite{Moch:2009hr}
do suggest  that the ${\cal O}(1/(1-x))$ part of the spacelike and timelike longitudinal evolution kernels might  actually be {\em identical} to any logarithmic accuracy.

\end{itemize}

\section{Conclusions}
\begin{itemize}
\item[$ \bullet$] Using a kinematically  modified \cite{Dokshitzer:2005bf}  physical evolution equation, evidence has been given that the {\em leading} threshold logarithms  at  {\em any}   eikonal order in the {\em momentum space} DIS and SIA {\em non-singlet} physical evolution kernels can be expressed in term of the {\em one loop} cusp anomalous dimension $A_1$, which
represents the {\em first step} towards threshold resummation {\em beyond} the leading eikonal level.
This result also explains the observed {\em universality} \cite{Grunberg:2009yi,Moch:2009hr}  of the {\em leading} logarithmic contributions to the physical kernels of the various non-singlet structure functions at {\em any} order \cite{Moch:2009hr}  in $1-x$.

\item[$ \bullet$] The present approach {\em does not} work for  {\em subleading} next-to-eikonal logarithms. 
However, there are hints  of the possible existence of an underlying (yet to be understood) Gribov-Lipatov like relation  in the special case of the $F_1$ DIS structure function and the corresponding $ F_T$ SIA transverse fragmentation function.
 
\end{itemize}

\end{document}